\documentclass[12pt]{article}
\usepackage{amsmath}
\usepackage{amssymb}
\usepackage{amsfonts}
\usepackage{amscd}
\usepackage[all]{xy}

\newcommand{\level}{\thicklines\line(1,0){60}}
\newcommand{\ord}{\thicklines\line(1,0){2}}
\newcommand{\trait}{\line(0,1){2}}
\newcommand{\fleche}{\thicklines\vector(0,1){18}}

\newsavebox{\dash}

\savebox{\dash}(5,20)[b]{\multiput(0,0)(0,4){3}{\trait}}

\DeclareMathOperator{\degen}{deg}

\title{
New ladder operators for a rational extension of the harmonic oscillator and superintegrability of some 
two-dimensional systems}

\author{Ian Marquette$^{1,}$\thanks{Electronic address: i.marquette@uq.edu.au} \ and Christiane Quesne$^{2,}$\thanks{Electronic address: cquesne@ulb.ac.be}\\
{\small\sl $^1$ School of Mathematics and Physics, The University of Queensland,}\\
{\small \sl Brisbane, QLD 4072, Australia}\\
{\small\sl $^2$ Physique Nucl\'eaire Th\'eorique et Physique Math\'ematique, 
Universit\'e Libre de Bruxelles,} \\ 
{\small \sl Campus de la Plaine CP229, Boulevard~du Triomphe, B-1050
Brussels, Belgium}}
\date{ }
\begin{document}
\baselineskip=22pt plus 1pt minus 1pt
\maketitle

\begin{abstract}
New ladder operators are constructed for a rational extension of the harmonic oscillator associated with type III Hermite exceptional orthogonal polynomials and characterized by an even integer $m$. The eigenstates of the Hamiltonian separate into $m+1$ infinite-dimensional unitary irreducible representations of the corresponding polynomial Heisenberg algebra. These ladder operators are used to construct a higher-order integral of motion for two superintegrable two-dimensional systems separable in cartesian coordinates. The polynomial algebras of such systems provide for the first time an algebraic derivation of the whole spectrum through their finite-dimensional unitary irreducible representations.
\end{abstract}

\vspace{0.5cm}


 
\newpage
%
%
\section{INTRODUCTION}

In the vast domain of quantum nonlinear oscillators, those constructed by supersymmetric quantum mechanical (SUSYQM) techniques play an important role (see, e.g., Refs.~\cite{sukumar, junker98, fernandez99, fernandez05, bermudez} and references quoted therein). In particular, a model that appeared in the early 90s \cite{dubov92, dubov94} and was re-discovered several times later on (see, e.g., Refs.~\cite{junker97, gomez04, carinena, fellows}) is of special interest because its eigenstates can be written in terms of exceptional orthogonal polynomials (EOP), a field that has received a lot of attention during the last few years \cite{gomez09, cq08, cq09, odake09, grandati11b, ho11b, gomez12, odake11, cq11, grandati12}. The polynomials involved in such a problem are indeed type III Hermite EOP \cite{grandati11a, ho11a}.\par
%
%
Ladder operators for harmonic oscillator supersymmetric partners are usually constructed  by combining the oscillator creation and annihilation operators with the supercharges \cite{fernandez99, fernandez05, junker97} or combinations of the latter \cite{bermudez, carballo, mateo}. Together with the Hamiltonian, such operators close a polynomial Heisenberg algebra (PHA), which may have infinite-dimensional, as well as finite-dimensional unitary irreducible representations (unirreps) \cite{fernandez99, carballo}. This construction was carried out for the rationally-extended oscillator model referred to above and the corresponding PHA was shown to have two unirreps, an infinite-dimensional one spanned by all excited states and a one-dimensional unirrep spanned by the ground state \cite{junker97}.\par
%
%
Apart from their own interest, ladder operators are also very useful in other contexts such as nuclear physics, quantum chemistry or condensed matter. They have also applications in the context of mathematical physics and more particularly in the field of superintegrable higher-dimensional systems. Considering more specifically the case of two-dimensional Hamiltonians, when one leaves the realm of well-studied quadratically superintegrable ones, i.e., those allowing two second-order integrals of motion (see, e.g., Refs.~\cite{winternitz, kalnins03, kalnins06, daska, ballesteros} and references quoted therein), the direct approach for determining the integrals of motion becomes more and more difficult as the order of the integrals increases. This is clearly shown in recent works on systems with third-order integrals of motion \cite{gravel02, gravel04, marquette09a, marquette09b}. For such a reason, some other approaches, based on ladder operators \cite{marquette10}, recurrence relations \cite{kalnins11}, or SUSYQM \cite{demircioglu, marquette09c}, have been proposed.\par
%
%
In two recent studies, the recurrence relation \cite{post} and the ladder operator \cite{marquette13a} methods have been applied to construct new superintegrable systems connected with EOP families. In the latter work, in particular, some two-dimensional systems related to type III Hermite EOP, as well as to type I, II, or III Laguerre EOP, were analyzed by means of standard ladder operators constructed by supersymmetric techniques. If the results proved entirely satisfactory for type I or II, this was not the case for type III because it was then not possible to derive the whole energy spectrum from the representations of the polynomial algebra generated by the integrals of motion.\par
%
%
The purpose of the present paper is to show that an adequate approach to the superintegrable systems connected with type III EOP may also be found provided some novel ladder operators are constructed for the constituent one-dimensional Hamiltonians. Here we plan to consider more specifically the two superintegrable systems built in Ref.~\cite{marquette13a} from the above-mentioned rationally-extended harmonic oscillator related to type III Hermite EOP.\par
%
%
In Sec.~II, some new ladder operators are constructed for such a nonlinear oscillator and their action on the Hamiltonian eigenstates is determined. In Sec.~III, they are applied to construct integrals of motion for the two superintegrable two-dimensional systems considered in Ref.~\cite{marquette13a}. The polynomial algebras generated by such integrals of motion are then shown to lead to an algebraic derivation of the spectra. Finally, Sec.~IV contains the conclusion.\par
%
%
\section{NEW LADDER OPERATORS FOR A RATIONALLY-EXTENDED HARMONIC OSCILLATOR}

Let
\begin{equation}
  H^{(\pm)} = - \frac{d^2}{dx^2} + V^{(\pm)}(x) - E_m, \quad V^{(\pm)}(x) = W^2(x) \mp W'(x) + E_m,
  \quad W(x) = - \bigl(\phi_m(x)\bigr)'
\end{equation}
be a pair of partner Hamiltonians in first-order SUSYQM \cite{cooper}, where $V^{(+)}(x) = x^2$ ($-\infty < x < \infty$) is the harmonic oscillator potential, while the factorization function and energy (such that $H^{(+)} \phi_m = 0$) are chosen as $\phi_m(x) = {\cal H}_m(x) \exp(x^2/2)$, $E_m = - 2m - 1$, with $m$ even \cite{fellows}. Here ${\cal H}_m(x)$ is a pseudo-Hermite polynomial, defined by ${\cal H}_m(x) = (- {\rm i})^m H_m({\rm i}x)$ in terms of a standard Hermite one. The two Hamiltonians intertwine with
\begin{equation}
  A = \frac{d}{dx} + W(x), \qquad A^{\dagger} = - \frac{d}{dx} + W(x), \qquad W(x) = - x - 
  \frac{{\cal H}'_m}{{\cal H}_m}, 
\end{equation}
as $A H^{(+)} = H^{(-)} A$, $A^{\dagger} H^{(-)} = H^{(+)} A^{\dagger}$, and the partner potential
\begin{equation}
  V^{(-)}(x) = x^2 - 2 \left[\frac{{\cal H}''_m}{{\cal H}_m} - \left(\frac{{\cal H}'_m}{{\cal H}_m}\right)^2 + 1
  \right]
\end{equation}
is a rationally-extended harmonic oscillator considered in many works \cite{dubov92, dubov94, junker97, gomez04, carinena, fellows, grandati11a, ho11a}.\par
%
%
Since, for even $m$, ${\cal H}_m(x)$ is strictly positive on the whole real line, the partner $H^{(-)}$ has an extra bound state below the oscillator spectrum, whose eigenfunction is proportional to $\phi_m^{-1}(x)$. As a consequence, the bound-state energies and wavefunctions of the two partners are given by \cite{fellows, marquette13a, marquette13b}
\begin{equation}
\begin{split}
  & E^{(+)}_{\nu} = 2 (\nu + m +1), \qquad \nu = 0, 1, 2, \ldots, \\
  & E^{(-)}_{\nu} = 2 (\nu + m +1), \qquad \nu = -m-1, 0, 1, 2, \ldots,
\end{split}  \label{eq:energy}
\end{equation}
and
\begin{equation}
\begin{split}
  & \psi^{(+)}_{\nu}(x) = {\cal N}^{(+)}_{\nu} H_{\nu}(x) e^{- \frac{1}{2}x^2}, \qquad \nu=0, 1, 2, \ldots, \\
  & \psi^{(-)}_{\nu}(x) = {\cal N}^{(-)}_{\nu} \frac{e^{- \frac{1}{2}x^2}}{{\cal H}_m(x)} y^{(m)}_{\nu+m+1}(x),
      \qquad \nu=-m-1, 0, 1, 2, \ldots, 
\end{split} \label{eq:psi}
\end{equation}
with
\begin{equation}
  y^{(m)}_0(x) = 1, \qquad y^{(m)}_{\nu+m+1}(x) = - {\cal H}_m(x) H_{\nu+1}(x) - 2m {\cal H}_{m-1}(x)
  H_{\nu}(x), \quad \nu=0, 1, 2, \ldots,
\end{equation}
and
\begin{equation}
\begin{split}
  & {\cal N}^{(+)}_{\nu} = \bigl(\sqrt{\pi} 2^{\nu} \nu!\bigr)^{-1/2}, \qquad \nu=0, 1, 2, \ldots, \\
  & {\cal N}^{(-)}_{-m-1} = \left(\frac{2^m m!}{\sqrt{\pi}}\right)^{1/2}, \qquad {\cal N}^{(-)}_{\nu} = 
       \bigl[\sqrt{\pi} 2^{\nu+1} (\nu+m+1) \nu!\bigr]^{-1/2}, \quad \nu=0, 1, 2, \ldots.  
\end{split} \label{eq:norm}
\end{equation}
The polynomials $y^{(m)}_n(x)$, $n=\nu+m+1$, $\nu=-m-1$, 0, 1, 2,~\ldots, are known as the (type III) Hermite EOP \cite{grandati11a, ho11a}. They form an orthogonal and complete set with respect to the positive-definite measure $\exp(-x^2) \bigl({\cal H}_m(x)\bigr)^{-2} dx$.\par
%
%
In Ref.~\cite{marquette13a}, we considered ladder operators $b^{\dagger} = A a^{\dagger} A^{\dagger}$, $b = A a A^{\dagger}$ for $H^{(-)}$ obtained from the creation and annihilation operators $a^{\dagger} = - d/dx + x$, $a = d/dx + x$, valid for $H^{(+)}$, and the supercharge operators $A^{\dagger}$, $A$ in the standard way \cite{fernandez99, fernandez05, junker97}. Here we plan to build some alternative ladder operators $c^{\dagger}$, $c$.\par
%
%
{}For such a purpose, let us first show that one can go from $H^{(+)}$ to $H^{(-)}$ (up to some additive constant) by another path using $m$ first-order SUSYQM transformations characterized by the supercharges
\begin{equation}
  \hat{A}_i = \frac{d}{dx} + \hat{W}_i(x), \quad \hat{A}_i^{\dagger} = - \frac{d}{dx} + \hat{W}_i(x), \quad 
  \hat{W}_i(x) = x + \frac{{\cal H}'_{i-1}}{{\cal H}_{i-1}} - \frac{{\cal H}'_i}{{\cal H}_i}, \quad i=1, 2, \ldots, m. 
\end{equation}
On defining 
\begin{equation}
  \hat{H}_i = - \frac{d^2}{dx^2} + x^2 - 2 \left[\frac{{\cal H}''_{i-1}}{{\cal H}_{i-1}} - \left(\frac{{\cal H}'_{i-1}}
  {{\cal H}_{i-1}}\right)^2\right] - 3, \qquad i=1, 2, \ldots, m+1,
\end{equation}
we indeed get $\hat{A}_i^{\dagger} \hat{A}_i = \hat{H}_i$ and $\hat{A}_i \hat{A}_i^{\dagger} = \hat{H}_{i+1} + 2$ for $i=1$, 2, \ldots,~$m$, which implies that $\hat{A}_i \hat{H}_i = \bigl(\hat{H}_{i+1} + 2\bigr) \hat{A}_i$ and $\hat{A}_i^{\dagger} \bigl(\hat{H}_{i+1} + 2\bigr) = \hat{H}_i \hat{A}_i^{\dagger}$. Since
\begin{equation}
  H^{(+)} = \hat{H}_1 + 2m + 4, \qquad H^{(-)} = \hat{H}_{m+1} + 2m + 2,
\end{equation}
we infer that
\begin{equation}
\begin{split}
  & \hat{A}_m \cdots \hat{A}_2 \hat{A}_1 H^{(+)} = (H^{(-)} + 2m + 2) \hat{A}_m \cdots \hat{A}_2 \hat{A}_1,
       \\ 
  & H^{(+)} \hat{A}^{\dagger}_1 \hat{A}^{\dagger}_2 \cdots \hat{A}^{\dagger}_m = \hat{A}^{\dagger}_1 
  \hat{A}^{\dagger}_2 \cdots \hat{A}^{\dagger}_m (H^{(-)} + 2m + 2),
\end{split}
\end{equation}
which proves the above assertion. It is worth observing here that the operators $\hat{A}_i$, $\hat{A}_i^{\dagger}$, $i=1$, 2,~\ldots, $m$, and $\hat{H}_i$, $i=2$, 3,~\ldots, $m$, are only auxiliary operators, some of which are singular at $x=0$. This means that we actually have a dressing chain of Hamiltonians \cite{mateo, veselov}.\par
%
%
This chain of $m$ first-order SUSYQM transformations from $H^{(+)}$ to $H^{(-)} + 2m + 2$ can be combined with the transformation from $H^{(-)}$ to $H^{(+)}$ to provide some raising and lowering operators for $H^{(-)}$,
\begin{equation}
  c^{\dagger} = A \hat{A}_1^{\dagger} \hat{A}_2^{\dagger} \cdots \hat{A}_m^{\dagger}, \qquad 
  c = \hat{A}_m \cdots \hat{A}_2 \hat{A}_1 A^{\dagger},
  \label{eq:c}
\end{equation}
which are ($m+1$)th-order differential operators. From the set of intertwining relations satisfied by $A^{\dagger}$ and $\hat{A}_i$, $i=1$, 2, \ldots,~$m$, it is indeed easy to prove that $c H^{(-)} = (H^{(-)}+2m+2) c$ or
\begin{equation}
  \xymatrixcolsep{5pc}\xymatrix@1{
  H^{(-)}  \ar@/_{10mm}/[rr]^{c}  \ar[r]^{A^{\dagger}} & H^{(+)}  \ar[r]^-{\hat{A}_m\cdots\hat{A}_{2}
  \hat{A}_{1}}  & H^{(-)}+2m+2}
\end{equation}
\par
%
%
The operators $H^{(-)}$, $c^{\dagger}$, and $c$ fulfil the commutation relations
\begin{equation}
\begin{split}
  & [H^{(-)}, c^{\dagger}] = (2m+2) c^{\dagger}, \qquad [H^{(-)}, c] = - (2m+2) c, \\
  & [c, c^{\dagger}] = Q(H^{(-)}+2m+2) - Q(H^{(-)}), 
\end{split}
\end{equation}
where
\begin{equation}
  Q(H^{(-)}) = H^{(-)} \prod_{i=1}^m (H^{(-)}-2m-2-2i)
\end{equation}
is a $(m+1)$th-order polynomial in $H^{(-)}$. They therefore form a PHA of $m$th order \cite{fernandez99, carballo}.\par
%
%
\begin{center}
\begin{picture}(100,100)(0,0)

\put(10,5){\thicklines\vector(0,1){90}}
\multiput(10,5)(0,30){3}{\ord}
\put(5,5){\makebox(4,0)[r]{\large 0}}
\put(5,35){\makebox(4,0)[r]{\large 10}}
\put(5,65){\makebox(4,0)[r]{\large 20}}
\put(5,90){\makebox(4,0)[r]{\large $E^{(-)}_{\nu}$}}

\put(20,5){\level}
\multiput(20,23)(0,6){10}{\level}
\put(30,82){\usebox{\dash}}
\put(50,82){\usebox{\dash}}
\put(70,82){\usebox{\dash}}

\multiput(30,5)(0,18){4}{\fleche}
\multiput(50,29)(0,18){2}{\fleche}
\put(50,65){\thicklines\line(0,1){12}}
\multiput(70,35)(0,18){2}{\fleche}
\put(70,71){\thicklines\line(0,1){6}}

\put(84,5){\makebox(4,0)[r]{\large -3}}
\put(84,23){\makebox(4,0)[r]{\large 0}}
\put(84,29){\makebox(4,0)[r]{\large 1}}
\put(84,35){\makebox(4,0)[r]{\large 2}}
\put(84,41){\makebox(4,0)[r]{\large 3}}
\put(84,47){\makebox(4,0)[r]{\large 4}}
\put(84,53){\makebox(4,0)[r]{\large 5}}
\put(84,59){\makebox(4,0)[r]{\large 6}}
\put(84,65){\makebox(4,0)[r]{\large 7}}
\put(84,71){\makebox(4,0)[r]{\large 8}}
\put(84,77){\makebox(4,0)[r]{\large 9}}

\end{picture}

\bigskip
\noindent
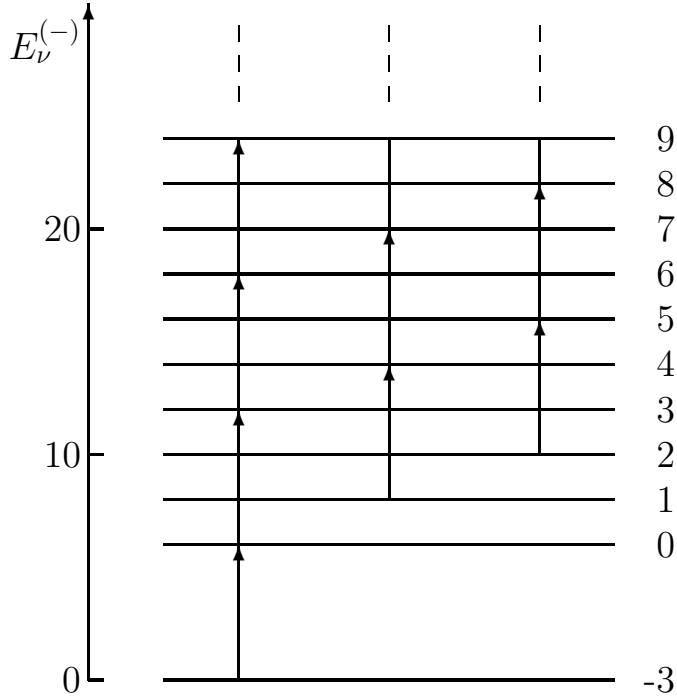
FIG.~1. Energy spectrum of $H^{(-)}$ and action of $c^{\dagger}$ on the eigenstates for $m=2$. The $\nu$ values are indicated on the right.

\end{center}
\par
%
%
The action of the raising operator $c^{\dagger}$ on the eigenfunctions $\psi^{(-)}_{\nu}(x)$ of $H^{(-)}$, defined  in (\ref{eq:psi})--(\ref{eq:norm}), can be easily calculated and is given by
\begin{equation}
\begin{split}
  & c^{\dagger} \psi^{(-)}_{-m-1} = \bigl[2^{m+1} (m+1)!\bigr]^{1/2} \psi^{(-)}_0, \\
  & c^{\dagger} \psi^{(-)}_{\nu} = - \bigl[2^{m+1} (\nu+m) (\nu+m-1) \cdots (\nu+1) (\nu+2m+2)\bigr]^{1/2}
       \psi^{(-)}_{\nu+m+1}, \\
  & \qquad \nu=0, 1, 2, \ldots.
\end{split} \label{eq:ladder-action}
\end{equation}
For that of $c$, we get
\begin{equation}
  c \psi^{(-)}_{\nu} = 0, \qquad \nu=-m-1, 1, 2, \ldots, m,
\end{equation}
as well as the Hermitian conjugate of Eq.~(\ref{eq:ladder-action}). We conclude that the PHA generated by $H^{(-)}$, $c^{\dagger}$, and $c$ has $m+1$ infinite-dimensional unirreps spanned by the states $\{\psi^{(-)}_{i+(m+1) j} \mid j=0,1,2,\ldots\}$ with $i=-m-1$, 1, 2, \ldots, $m$, respectively.\par
%
%
The spectrum of $H^{(-)}$ and the action of $c^{\dagger}$ on the eigenstates is displayed in Fig.~1 for the $m=2$ example, in which case the PHA has three infinite-dimensional unirreps.\par
%
%
\section{APPLICATION OF THE NEW LADDER OPERATORS TO SOME SUPERINTEGRABLE TWO-DIMENSIONAL SYSTEMS}
\setcounter{equation}{0}

Let us consider some two-dimensional Hamiltonians allowing separation of variables in cartesian coordinates,
\begin{equation}
  H = H_x + H_y = - \frac{d^2}{dx^2} - \frac{d^2}{dy^2} + V_x(x) + V_y(y),  \label{eq:H}
\end{equation}
and let us assume that there exist ladder operators $(a_x^{\dagger}, a_x)$ and $(a_y^{\dagger}, a_y)$ in both axes that are differential operators of order $k_1$ and $k_2$, respectively, and satisfy the defining relations of two PHA's,
\begin{equation}
\begin{split}
  & [H_x, a_x^{\dagger}] = \lambda_x a_x^{\dagger}, \qquad [H_x, a_x] = - \lambda_x a_x, \qquad
      [a_x, a_x^{\dagger}] = Q(H_x + \lambda_x) - Q(H_x), \\
  & [H_y, a_y^{\dagger}] = \lambda_y a_y^{\dagger}, \qquad [H_y, a_y] = - \lambda_y a_y, \qquad
      [a_y, a_y^{\dagger}] = S(H_y + \lambda_y) - S(H_y).      
\end{split}  \label{eq:PHA}
\end{equation}
Here $\lambda_x$ and $\lambda_y$ are constants, while $Q(H_x)$ and $S(H_y)$ are polynomials.\par
%
%
The separation of variables in cartesian coordinates implies the existence of a second-order integral of motion $H_x - H_y$, showing that the two-dimensional system (\ref{eq:H}) is integrable. From the ladder operators, one can construct additional polynomial operators commuting with $H$, $a_x^{\dagger n_1} a_y^{n_2}$ and $a_x^{n_1} a_y^{\dagger n_2}$, where $n_1$, $n_2 \in \mathbb{Z}^{+}$ are chosen such that $n_1 \lambda_x = n_2 \lambda_y = \lambda$ \cite{marquette10}. Hence system (\ref{eq:H}) possesses three algebraically independent integrals of motion and is superintegrable. It is worth stressing that from ladder operators of rather low order $k_1$, $k_2$, one generates integrals of motion of higher order $k_1 n_1 + k_2 n_2$ in a nice factorized form that would be difficult to obtain in a direct approach.\par
%
%
The integrals of motion
\begin{equation}
  K = \frac{1}{2\lambda}(H_x-H_y),\qquad I_+ = a_x^{\dagger n_1} a_y^{n_2}, \qquad I_- = a_x^{n_1}
  a_y^{\dagger n_2}
\end{equation}
generate the polynomial algebra of the system
\begin{equation}
\begin{split}
  & [K,I_{\pm}] = \pm I_{\pm}, \qquad [I_-,I_+] = F_{n_1,n_2}(K+1,H) - F_{n_1,n_2}(K,H), \\ 
  & F_{n_1,n_2}(K,H) = \prod_{i=1}^{n_1} Q\left(\frac{H}{2}+\lambda K-(n_1-i)\lambda_{x}\right)
       \prod_{j=1}^{n_2}S\left(\frac{H}{2}-\lambda K+j\lambda_y\right),
\end{split} \label{eq:algebra}
\end{equation}
which is of order $k_1n_1+k_2n_2-1$. Such a polynomial algebra is a deformed $u(2)$ algebra and its finite-dimensional unirreps can be found by realizing it as a generalized deformed oscillator algebra $\{b^t, b, N\}$. The operators $b^t=I_+$, $b=I_-$, $N=K-u$ and $\Phi(H,u,N) = F_{n_1,n_2}(K,H)$ indeed satisfy the defining relations of such an algebra \cite{bonatsos},
\begin{equation}
  [N,b^t] = b^t , \qquad [N,b] = -b, \qquad b^t b = \Phi(H,u,N), \qquad bb^t = \Phi(H,u,N+1), 
\end{equation}
where $u$ is some constant and $\Phi(H,u,N)$ is called ``structure function''. If the latter satisfies the properties
\begin{equation}
  \Phi(E,u,0) = 0, \qquad \Phi(E,u,p+1) = 0, \qquad \Phi(E,u,n) >0, \quad  n=1,2,\ldots,p,  \label{eq:constraints} 
\end{equation}
then the deformed oscillator algebra has an energy-dependent Fock space of dimension $p+1$ with a Fock basis $|E,n\rangle$, $n=0$, 1,~\ldots, $p$, fulfilling
\begin{equation}
\begin{split}
  & H |E,n\rangle = E |E,n\rangle, \qquad N |E,n\rangle = n |E,n\rangle, \qquad b |E,0\rangle = 0, \qquad 
      b^t |E,p\rangle = 0, \\
  & b^t |E,n\rangle = \sqrt{\Phi(E,u,n+1)} |E,n+1\rangle, \qquad b |E,n\rangle = \sqrt{\Phi(E,u,n)} |E,n-1\rangle. 
\end{split}
\end{equation}
These relations can be used to obtain the ($p+1$)-dimensional unirreps of the polynomial algebra (\ref{eq:algebra}) and the corresponding degenerate energy spectrum of the system.\par
%
%
Let us illustrate this method on two systems already considered in Ref.~\cite{marquette13a} by taking advantage this time of the new ladder operators introduced in Sec.~II.\par
%
%
\subsection{Combination of a rationally-extended oscillator with a standard one}

Let us consider the two-dimensional system given by (\ref{eq:H}) with respectively in the $x$-axis the superpartner of the harmonic oscillator considered in Sec.~II and in the $y$-axis the harmonic oscillator itself,
\begin{equation}
\begin{split}
  & H_x = H^{(-)} - 2m - 1 = - \frac{d^2}{dx^2} + x^2 - 2 \left[\frac{{\cal H}''_m}{{\cal H}_m} - 
       \left(\frac{{\cal H}'_m}{{\cal H}_m}\right)^2 + 1\right], \qquad \text{$m$ even}, \\
  & H_y = - \frac{d^2}{dy^2} + y^2.
\end{split}  \label{eq:H-case1}
\end{equation}
On taking for $(a_x^{\dagger}, a_x)$ the operators $(c^{\dagger}, c)$, defined in (\ref{eq:c}), and for $(a_y^{\dagger}, a_y)$ standard harmonic oscillator creation and annihilation operators, it is obvious that Eq.~(\ref{eq:PHA}) is satisfied with $\lambda_x = 2m+2$, $\lambda_y = 2$, and
\begin{equation}
  Q(H_x) = (H_x+2m+1) \prod_{i=1}^m (H_x-1-2i), \qquad S(H_y) = H_y-1.  \label{eq:Q}
\end{equation}
The structure function $\Phi(E,u,x)$ is obtained from Eqs.~(\ref{eq:algebra}) (with $n_1 = 1$, $n_2 = m+1$, $\lambda = 2(m+1)$) and (\ref{eq:Q}) as
\begin{equation}
\begin{split}
  & \Phi(E,u,x) \\
  &= \left(\frac{E}{2} + 2(m+1)(x+u) + 2m + 1\right) \prod_{i=1}^m \left(\frac{E}{2} + 2(m+1)(x+u)
      - 1 - 2i\right) \\
  & \quad \times \prod_{j=1}^{m+1} \left(\frac{E}{2} - 2(m+1)(x+u) + 2j - 1\right).
\end{split}
\end{equation}
\par
%
%
{}From this structure function and the first constraint of Eq.~(\ref{eq:constraints}), we get three different types of solutions for the parameter $u$,
\begin{equation}
\begin{split}
  & 2(m+1) u_1 = - \frac{E}{2} - 2m - 1, \\
  & 2(m+1) u_2 = - \frac{E}{2} + 2l + 1, \qquad l \in \{1, 2, \ldots, m\}, \\
  & 2(m+1) u_3 = \frac{E}{2} + 2l - 1, \qquad l \in \{1, 2, \ldots, m+1\}.
\end{split}
\end{equation}
The finite-dimensional unirreps are calculated from the two other constraints of Eq.~(\ref{eq:constraints}) and come from the first two types of solutions $u_1$ and $u_2$. They are associated with the energies
\begin{equation}
  E_1 = 2 [(m+1)p +1 - k],  \label{eq:E-1}
\end{equation}
\begin{equation}
  E_2 = 2 [(m+1)(p+1) + l - k + 1],  \label{eq:E-2}
\end{equation}
and with the structure functions
\begin{equation}
\begin{split}
  \Phi_1 &= 2^{2(m+1)} (m+1) x \prod_{i=1}^m [(m+1)x - m - 1 - i] \\
  & \quad \times \prod_{j=1}^{m+1} [(m+1)(p+1-x) - m + j - k],
\end{split}
\end{equation}
\begin{equation}
\begin{split}
  \Phi_2 &= 2^{2(m+1)} [(m+1) x + m + 1 + l] \prod_{i=1}^m [(m+1)x + l - i] \\
  & \quad \times \prod_{j=1}^{m+1} [(m+1)(p+1-x) + j - k].
\end{split}
\end{equation}
Here $p \in \mathbb{N}$, $k \in \{1, 2, \ldots, m+1\}$, and $l \in \{1, 2, \ldots, m\}$. We conclude that there are altogether $(m+1)^2$ unirreps characterized by the same $p \in \mathbb{N}$.\par
%
%
Let us now show that, in contrast with the previous approach \cite{marquette13a}, the present one provides all the levels of the physical energy spectrum with their corresponding degeneracy. From Eqs.~(\ref{eq:energy}) and (\ref{eq:H-case1}), the energy spectrum of $H$ is indeed obtained as
\begin{equation}
  E = E_x + E_y = 2(\nu_x + \nu_y + 1), \qquad \nu_x = -m-1, 0, 1, 2, \ldots, \qquad \nu_y = 0, 1, 2, \ldots.
\end{equation}
On setting
\begin{equation}
  E_N = 2N, \qquad N=\nu_x+\nu_y+1,  \label{eq:E}
\end{equation}
we obtain
\begin{equation}
  \degen(E_N) = \begin{cases}
    1 & \text{if $N = -m, -m+1, \ldots , -1$},\\
    N+1 & \text{if $N = 0, 1, 2, \ldots$.} 
  \end{cases}  \label{eq:deg} 
\end{equation}
On defining then $\nu_x = (m+1) n_x + a_1$, $\nu_y = (m+1) n_y + a_2$, with $n_x, n_y \in \mathbb{N}$, $a_1 \in \{-m-1, 1, 2, \ldots, m\}$, and $a_2 \in \{0, 1, \ldots, m\}$, $E_N$, as given in Eq.~(\ref{eq:E}), can be rewritten as
\begin{equation}
  E_N = 2[(m+1)(n_x+n_y) + a_1 + a_2 + 1].  \label{eq:E-bis}
\end{equation}
It is then straightforward to see that $E_1$ and $E_2$, defined in (\ref{eq:E-1}) and (\ref{eq:E-2}), correspond to Eq.~(\ref{eq:E-bis}) with $n_x + n_y = p \in \mathbb{N}$, $a_2 = m+1-k \in \{0, 1, \ldots, m\}$, and $a_1 = -m-1$ or $a_1 = l \in \{1, 2, \ldots, m\}$, respectively.\par
%
%
In the simplest $m=2$ case, i.e., for Potential 1 in Ref.~\cite{marquette09a}, which is one of Gravel's systems \cite{gravel04}, the polynomial algebra (\ref{eq:algebra}) has nine unirreps for each $p \in \mathbb{N}$, associated with the energies $6p-4$, $6p-2$, $6p$, $6p+4$, $(6p+6)^2$, $(6p+8)^2$, and $6p+10$, respectively. On the other hand, the sequence of energy levels with their degeneracy is $-4$, $-2$, 0, $2^2$, $4^3$, $6^4$, $8^5$, $10^6$,~\ldots. Only the lowest ones belong to a single unirrep (for instance, $-4$ is obtained from $6p-4$ with $p=0$ and $p+1=1$), whereas the remaining ones belong to several unirreps (for instance, $10^6$ is obtained from $6p-2$ with $p=2$ and $p+1=3$, $6p+4$ with $p=1$ and $p+1=2$, and $6p+10$ with $p=0$ and $p+1=1$).\par
%
%
In the general case, i.e., for an arbitrary even value of $m$, a detailed analysis from the two solutions $E_{1}$ and $E_{2}$ enabled us to recover the degeneracies (\ref{eq:deg}). On using 
\begin{equation}
  N=(m+1)\lambda +\mu,
\end{equation}
we found the number of unirreps per level given in Table I, where we also list the corresponding set of $p$ values with their number of occurrences and the total degeneracy. \par
%
%
\begin{table}[h!]

\caption{Set of $p$ values with their number of occurrences, number $\cal N$ of unirreps per level, and total level degeneracy for the polynomial algebra (\ref{eq:algebra}) corresponding to Hamiltonian (\ref{eq:H}), (\ref{eq:H-case1}).}

\begin{center}
\begin{tabular}{lllll}
  \hline\hline\\[-0.2cm]
  $\lambda$ & $\mu$ & $p$ & $\cal N$ & $\degen(E_N)$\\[0.2cm]
  \hline\\[-0.2cm]
  $-1$ & $1, 2, \ldots, m$ & 0 & 1 & 1 \\[0.2cm]
  0 & 0 & 0 & 1 & 1 \\[0.2cm]
  0 & $1, 2, \ldots, m$ & 1 & $\mu$ & $N+1$  \\[0.2cm]
  & & $0^{\mu-1}$ & & \\[0.2cm]
  $1, 2, \ldots$ &0 & $\lambda$ & $m+1$ & $N+1$ \\[0.2cm]
  & & $(\lambda-1)^{m}$ & & \\[0.2cm]
  $1, 2, \ldots$& $1, 2, \ldots,m$ & $\lambda+1$ & $m+1$ & $N+1$ \\[0.2cm]
  & & $\lambda^{\mu-1}$ & & \\[0.2cm]
  & & $(\lambda-1)^{m-\mu+1}$ & & \\[0.2cm]
  \hline \hline
\end{tabular}
\end{center}

\end{table}
\par
%
%
\subsection{Combination of two rationally-extended oscillators}

Let us now consider the case where
\begin{equation}
\begin{split}
  & H_x = - \frac{d^2}{dx^2} + x^2 - 2 \left[\frac{{\cal H}''_{m_1}}{{\cal H}_{m_1}} - 
       \left(\frac{{\cal H}'_{m_1}}{{\cal H}_{m_1}}\right)^2 + 1\right], \\
  & H_y = - \frac{d^2}{dy^2} + y^2 - 2 \left[\frac{{\cal H}''_{m_2}}{{\cal H}_{m_2}} - 
       \left(\frac{{\cal H}'_{m_2}}{{\cal H}_{m_2}}\right)^2 + 1\right],
\end{split}  \label{eq:H-case2} 
\end{equation}
with $m_1$ and $m_2$ even and such that $m_1 \ge m_2$. This includes another system obtained by Gravel \cite{gravel04} (Potential 6 in Ref.~\cite{marquette09a}) for $m_1 = m_2 = 2$.\par
%
%
On taking for $(a_x^{\dagger}, a_x)$ and $(a_y^{\dagger}, a_y)$ operators of type (\ref{eq:c}) (with $m \to m_1$ and $m \to m_2$, respectively) and on noting that Eq.~(\ref{eq:PHA}) is satisfied with $\lambda_x = 2m_1+2$, $\lambda_y = 2m_2+2$, $n_1 = m_2+1$, $n_2 = m_1+1$, $\lambda = 2(m_1+1)(m_2+1)$, and with both $Q(H_x)$ and $S(H_y)$ assuming a form similar to $Q(H_x)$ in (\ref{eq:Q}) (with $m \to m_1$ and $m \to m_2$, respectively), we arrive at the following structure function: 
\begin{equation}
\begin{split}
  &\Phi(E,u,x) \\
  & = \prod_{i=1}^{m_2+1} \biggl[\biggl(\frac{E}{2} + 2(m_1+1)(m_2+1)(x+u) - (2m_1+2)(m_2+1-i)
       + 2m_1+1\biggr) \\
  & \quad \times \prod_{k=1}^{m_1} \biggl(\frac{E}{2} + 2(m_1+1)(m_2+1)(x+u) - (2m_1+2)(m_2+1-i)
       - 1-2k\biggr)\biggr] \\
  & \quad \times \prod_{j=1}^{m_1+1} \biggl[\biggl(\frac{E}{2} - 2(m_1+1)(m_2+1)(x+u) + (2m_2+2)j
       + 2m_2+1\biggr) \\
  & \quad \times \prod_{l=1}^{m_2} \biggl(\frac{E}{2} - 2(m_1+1)(m_2+1)(x+u) + (2m_2+2)j
       - 1-2l\biggr)\biggr]. 
\end{split}
\end{equation}
\par
%
%
On proceeding as in the previous case, it can be easily shown that among the four different types of solutions for the parameter $u$, only the first two
\begin{equation}
\begin{split}
  & 2(m_1+1)(m_2+1) u_1 = - \frac{E}{2} + (2m_1+2)(m_2+1-q)  - 2m_1 - 1, \\
  & 2(m_1+1)(m_2+1) u_2 = - \frac{E}{2} + (2m_1+2)(m_2+1-q)  + 2r + 1,  
\end{split}
\end{equation}
with $q \in \{1, 2, \ldots, m_2+1\}$ and $r \in \{1, 2, \ldots, m_1\}$, lead to appropriate finite-dimensional unirreps of the polynomial algebra (\ref{eq:algebra}). The resulting energies are
\begin{equation}
\begin{split}
  & E_{11} = 2 [(m_1+1)(m_2+1)(p+2) - (m_1+1)q - (m_2+1)s - (m_1+m_2+1)], \\
  & E_{12} = 2 [(m_1+1)(m_2+1)(p+2) - (m_1+1)q - (m_2+1)s - m_1 + t], \\
  & E_{21} = 2 [(m_1+1)(m_2+1)(p+2) - (m_1+1)q - (m_2+1)s - m_2 + r], \\
  & E_{22} = 2 [(m_1+1)(m_2+1)(p+2) - (m_1+1)q - (m_2+1)s + r + t + 1], 
\end{split}
\end{equation}
with $p \in \mathbb{N}$, $q \in \{1, 2, \ldots, m_2+1\}$, $r \in \{1, 2, \ldots, m_1\}$, $s \in \{1, 2, \ldots, m_1+1\}$, and $t \in \{1, 2, \ldots, m_2\}$. The corresponding structure functions are given by
\begin{equation}
\begin{split}
  \Phi_{11} & = 2^{2(m_1+1)(m_2+1)} \prod_{i=1}^{m_2+1} \Bigl[(m_1+1) \Bigl((m_2+1)x - q + i\Bigr) \\
  & \quad \times \prod_{k=1}^{m_1} \Bigl((m_1+1)(m_2+1)x + (m_1+1)(i-q-1) - k\Bigr)\Bigr] \\
  & \quad \times \prod_{j=1}^{m_1+1} \Bigl[(m_2+1) \Bigl((m_1+1)(p+1-x) + j - s\Bigr) \\
  & \quad \times \prod_{l=1}^{m_2} \Bigl((m_1+1)(m_2+1)(p+1-x) + (m_2+1)(j-s-1) - l\Bigr)\Bigr],
\end{split}
\end{equation}
\begin{equation}
\begin{split}
  \Phi_{12} & = 2^{2(m_1+1)(m_2+1)} \prod_{i=1}^{m_2+1} \Bigl[(m_1+1) \Bigl((m_2+1)x - q + i\Bigr) \\
  & \quad \times \prod_{k=1}^{m_1} \Bigl((m_1+1)(m_2+1)x + (m_1+1)(i-q-1) - k\Bigr)\Bigr] \\
  & \quad \times \prod_{j=1}^{m_1+1} \Bigl[\Bigl((m_1+1)(m_2+1)(p+1-x) + (m_2+1)(j-s+1) + t\Bigr) \\
  & \quad \times \prod_{l=1}^{m_2} \Bigl((m_1+1)(m_2+1)(p+1-x) + (m_2+1)(j-s) + t - l\Bigr)\Bigr],
\end{split}
\end{equation}
\begin{equation}
\begin{split}
  \Phi_{21} & = 2^{2(m_1+1)(m_2+1)} \prod_{i=1}^{m_2+1} \Bigl[\Bigl((m_1+1)(m_2+1)x + (m_1+1)(i-q+1)
       + r\Bigr) \\
  & \quad \times \prod_{k=1}^{m_1} \Bigl((m_1+1)(m_2+1)x + (m_1+1)(i-q) + r - k\Bigr)\Bigr] \\
  & \quad \times \prod_{j=1}^{m_1+1} \Bigl[(m_2+1) \Bigl((m_1+1)(p+1-x) + j - s\Bigr) \\
  & \quad \times \prod_{l=1}^{m_2} \Bigl((m_1+1)(m_2+1)(p+1-x) + (m_2+1)(j-s-1) - l\Bigr)\Bigr],
\end{split}
\end{equation}
\begin{equation}
\begin{split}
  \Phi_{22} & = 2^{2(m_1+1)(m_2+1)} \prod_{i=1}^{m_2+1} \Bigl[\Bigl((m_1+1)(m_2+1)x + (m_1+1)(i-q+1)
       + r \Bigr) \\
  & \quad \times \prod_{k=1}^{m_1} \Bigl((m_1+1)(m_2+1)x + (m_1+1)(i-q) + r - k\Bigr)\Bigr] \\
  & \quad \times \prod_{j=1}^{m_1+1} \Bigl[\Bigl((m_1+1)(m_2+1)(p+1-x) + (m_2+1)(j-s+1) + t\Bigr) \\
  & \quad \times \prod_{l=1}^{m_2} \Bigl((m_1+1)(m_2+1)(p+1-x) + (m_2+1)(j-s) + t - l\Bigr)\Bigr],
\end{split}
\end{equation}
respectively. This time, there are altogether $(m_1+1)^2 (m_2+1)^2$ unirreps characterized by the same $p \in \mathbb{N}$.\par
%
%
On the other hand, from Sec.~II we know that the physical energy spectrum is given by
\begin{equation}
  E = E_x + E_y = 2(\nu_x + \nu_y + 1), \quad \nu_x = -m_1-1, 0, 1, 2, \ldots, \quad \nu_y = -m_2-1, 0, 1, 2,
      \ldots.
\end{equation}
With an equation similar to Eq.~(\ref{eq:E}), this leads to the degeneracies
\begin{equation}
  \degen(E_N) = \begin{cases}
    1 & \text{if $N = -2m-1$}, \\
    2 & \text{if $N = -m, -m+1, \ldots , -1$},\\
    N+2 & \text{if $N = 0, 1, 2, \ldots$.} 
  \end{cases}  \label{eq:deg-bis} 
\end{equation}
On setting now $\nu_x = (m_1+1) [(m_2+1)n_x + a_3] + a_1$, $\nu_y = (m_2+1) [(m_1+1)n_y + a_4] + a_2$, with $n_x, n_y \in \mathbb{N}$, $a_1 \in \{-m_1-1, 1, 2, \ldots, m_1\}$, $a_2 \in \{-m_2-1, 1, 2, \ldots, m_2\}$, $a_3 \in \{0, 1, \ldots, m_2\}$, $a_4 \in \{0, 1, \ldots, m_1\}$, $E_N$ can be rewritten as 
\begin{equation}
  E_N = 2 [(m_1+1)(m_2+1) (n_x+n_y) + (m_1+1) a_3 + (m_2+1) a_4 + a_1 + a_2 + 1].
\end{equation}
We then immediately see that $E_{11}$, $E_{12}$, $E_{21}$, and $E_{22}$ correspond to $E_N$ with $n_x+n_y = p \in \mathbb{N}$, $a_3 = m_2+1-q \in \{0, 1, \ldots, m_2\}$, $a_4 = m_1+1-s \in \{0, 1, \ldots, m_1\}$, and $a_1 = -m_1-1$, $a_2 = -m_2-1$, or $a_1 = -m_1-1$, $a_2 = t \in \{1, 2, \ldots, m_2\}$, or $a_1 = r \in \{1, 2, \ldots, m_1\}$, $a_2 = -m_2-1$, or $a_1 = r \in \{1, 2, \ldots, m_1\}$, $a_2 = t \in \{1, 2, \ldots, m_2\}$, respectively. The polynomial algebra of the system therefore provides the whole energy spectrum.\par
%
%
We checked on several examples that it also accounts for the level degeneracies (\ref{eq:deg-bis}), which are in general obtained through the use of several unirreps. For $m_1 = m_2 = m$, for instance, we got the number of unirreps per level given in Table II, where
\begin{equation}
  N = \lambda (m+1)^2 + \mu, \qquad \mu = \rho (m+1) + \sigma.
\end{equation}
\par
%
%
\begin{table}[h!]

\caption{Set of $p$ values with their number of occurrences, number $\cal N$ of unirreps per level, and total level degeneracy for the polynomial algebra (\ref{eq:algebra}) corresponding to Hamiltonian (\ref{eq:H}), (\ref{eq:H-case2}) in the $m_1 = m_2 = m$ case.}

\begin{center}
\begin{tabular}{llllll}
  \hline\hline\\[-0.2cm]
  $\lambda$ & $\rho$ & $\sigma$ & $p$ & $\cal N$ & $\degen(E_N)$\\[0.2cm]
  \hline\\[-0.2cm]
  $-1$ & $m-1$ & 1 & 0 & 1 & 1 \\[0.2cm]
  $-1$ & $m$ & $1, 2, \ldots, m$ & $0^2$ & 2 & 2 \\[0.2cm]
  0 & $m$ & $1, 2, \ldots, m$ & $1^2$ & $\mu$ & $N+2$ \\[0.2cm]
  &&& $0^{\mu-2}$ && \\[0.2cm]
  0 & $m-1$ & 1 & 1 & $\mu+1$ & $N+2$ \\[0.2cm]
  &&& $0^{\mu}$ && \\[0.2cm]
  0 & $0, 1, \ldots, m$ & 0 & $0^{\mu+2}$ & $\mu+2$ & $N+2$ \\[0.2cm]
  0 & $0, 1, \ldots, m-2$ & 1 & $0^{\mu+2}$ & $\mu+2$ & $N+2$ \\[0.2cm]
  0 & $0, 1, \ldots, m-1$ & $2, 3, \ldots, m$ & $0^{\mu+2}$ & $\mu+2$ & $N+2$ \\[0.2cm]
  $1, 2, \ldots$ & $0, 1, \ldots,m$ & 0 & $\lambda^{\mu+2}$ & $(m+1)^2$ & $N+2$ \\[0.2cm]
  &&& $(\lambda-1)^{(m+1)^2-\mu-2}$ && \\[0.2cm]
  $1, 2, \ldots$ & $0, 1, \ldots,m-2$ & 1 & $\lambda^{\mu+2}$ & $(m+1)^2$ & $N+2$ \\[0.2cm]
  &&& $(\lambda-1)^{(m+1)^2-\mu-2}$ && \\[0.2cm]
  $1, 2, \ldots$ & $0, 1, \ldots,m-1$ & $2, 3, \ldots, m$ & $\lambda^{\mu+2}$ & $(m+1)^2$ & $N+2$ \\[0.2cm]
  &&& $(\lambda-1)^{(m+1)^2-\mu-2}$ && \\[0.2cm]
  $1, 2, \ldots$ & $m-1$ & 1 & $\lambda+1$ & $(m+1)^2$ & $N+2$ \\[0.2cm]
  &&& $\lambda^{\mu}$ && \\[0.2cm]
  &&& $(\lambda-1)^{(m+1)^2-\mu-1}$ && \\[0.2cm]
  $1, 2, \ldots$ & $m$ & $1, 2, \ldots, m$ & $(\lambda+1)^2$ & $(m+1)^2$ & $N+2$ \\[0.2cm]
  &&& $\lambda^{\mu-2}$ && \\[0.2cm]
  &&& $(\lambda-1)^{(m+1)^2-\mu}$ && \\[0.2cm]
  \hline \hline
\end{tabular}
\end{center}

\end{table}
\par
%
%
Here we have taken the convenient and uniform choice $n_{1}=m_{2}+1$ and $n_{2}=m_{1}+1$. However it is worth observing that whenever $m_1+1$ and $m_2+1$ have a common factor, i.e., $m_1+1 = \mu \nu_1$ and $m_2+1 = \mu \nu_2$, there exists a simpler choice for $n_1$ and $n_2$, namely $n_1 = \nu_2$ and $n_2 = \nu_1$, which would lead to a lower-order polynomial algebra. In any case, it is well known that if ladder operators provide an easy method for constructing integrals of motion, the resulting algebraic structures are not necessarily the simplest ones that can be obtained \cite{marquette10}. 
\par
%
%
\section{CONCLUSION}

In the present work, the construction of ladder operators for a well-known rational extension of the harmonic oscillator, associated with type III Hermite EOP and characterized by an even integer $m$, has been reconsidered. Novel operators closing a PHA of $m$th order have been built and it has been shown that the eigenstates of this rational extension separate into $m+1$ infinite-dimensional unirreps of the PHA.\par
%
%
Such ladder operators have then been applied to construct a higher-order integral of motion for two superintegrable two-dimensional systems separable in cartesian coordinates. It has been proved that the polynomial algebras of these systems provide an algebraic derivation of the whole energy spectrum through their $(p+1)$-dimensional unirreps. The degeneracy of the energy levels in general results from the union of several unirreps.\par
%
%
In conclusion, we have shown that as it was already the case for superintegrable systems connected with type I or II EOP, a full algebraic treatment may also be found for those related to type III ones provided some appropriate ladder operators are constructed.\par
%
%
The integrals we constructed with these new ladder operators are of higher order than the ones generated by standard ladder operators. These results point out that beyond quadratically superintegrable systems the lowest-order integrals do not necessarily provide the whole energy spectrum with its degeneracies and that to get the latter one might need integrals of higher order.\par
%
%
In a future work, we hope to be able to carry out a similar study for the systems \cite{marquette13a} built from extended radial oscillators \cite{grandati11b, ho11b}, as well as for those that might be built from other extended potentials \cite{marquette13b, odake13a, odake13b}.\par
%
%
\section*{ACKNOWLEDGMENTS}

The research of I.\ M.\ was supported by the Australian Research Council through Discovery Project DP110101414 and Discovery Early Career Researcher Award DE130101067.
\par
%
%

\end{document}